\documentclass{PoS}

\usepackage{amsmath}
\usepackage{amsthm}

\title{Quark number susceptibilities at finite chemical potential from fugacity expansion }

\ShortTitle{Quark number susceptibilities at finite chemical potential from fugacity expansion }

\author{\speaker{Hans-Peter Schadler}\\
        Institut f\"ur Physik, Karl-Franzens Universit\"at Graz, 8010 Graz, Austria\\
        E-mail: \email{hps@abyle.org}}
        
\author{Christof Gattringer\\
        Institut f\"ur Physik, Karl-Franzens Universit\"at Graz, 8010 Graz, Austria\\
        E-mail: \email{christof.gattringer@uni-graz.at}}

\abstract{Generalized quark number susceptibilities are expected to be good probes for the phase transitions in QCD and 
the search of a possible critical point. However, their computation in lattice QCD is plagued by the complex action problem which appears at 
finite chemical potential $\mu$. In this work we explore the possibilities of an expansion in the fugacity parameter $e^{\mu\beta}$ 
which has features that make, in particular quark number related bulk observables easily accessible. We present results at finite chemical 
potential for generalized susceptibilities up to the 4th order as well as their ratios and compare them to model calculations.}

\FullConference{The 32nd International Symposium on Lattice Field Theory\\
		 23-28 June, 2014\\
		 Columbia University New York, NY}

\begin{document}

\section{Motivation}\label{sec:one}
It has been proposed that fluctuations of conserved charges like the baryon number, i.e., derivatives of the partition sum with respect to chemical potential, 
may be good probes for the QCD phase transition lines and the search of a possible critical point. Furthermore they can be used for the extraction of the 
freeze-out temperature in heavy ion collisions~\cite{fluctuationsprobes}. Not only are these observables accessible in experiment, but their ratios are also 
independent of the interaction/fireball volume of the collisions, which makes them interesting quantities to study.

\hspace*{10mm}
While we can easily calculate observables at finite temperature using lattice QCD, we are still plagued by the sign problem when going to finite baryon 
chemical potential. Therefore, also when it is clear what observables we should calculate, it is challenging to actually 
perform these calculations and extract reliable 
values for the cumulants, especially in the important regime of large chemical potential and low temperature. 

\hspace*{10mm}
Here we report on testing the fugacity 
expansion as a method to calculate these quantities at finite chemical potential and present results for the quark number density, for higher derivatives and 
especially ratios of derivatives using Wilson and staggered fermions in full dynamical lattice QCD. 
Recent results on the fugacity expansion have shown that in some cases it can 
have better convergence properties than a Taylor expansion~\cite{Z3fugacity}, which would make this approach an interesting alternative. 
Compared to previous studies~\cite{oldfugacity, domdecomp} we are now able to go to larger lattices and smaller quark masses with reasonable 
statistics and accuracy. This should give a better estimate of the quality of this expansion technique for QCD.

\section{Quark number related observables from fugacity expansion}\label{sec:two}

We start with the definition of the fugacity series, which provides the relation between the grand canonical fermion determinant $\det[D(\mu)]$ for a given 
chemical potential $\mu\beta$ ($\beta = 1/T$, $k_B=1$) and the canonical determinants $D^{(q)}$ with fixed net quark number $q$:
	\begin{equation}\label{eq:fug1}
		\det[D(\mu)] \; = \; \sum_{q=-q_{\text{cut}}}^{q_{\text{cut}}} e^{\,\mu \beta q}\,D^{\,(q)} \; .
	\end{equation}
The canonical determinants can be computed with the Fourier integral
\begin{equation}\label{eq:fug2}
	D^{(q)} \; = \; \frac{1}{2\pi}\int_{-\pi}^{\pi}d\phi\,e^{-iq\phi}\,\det[D(\mu \beta = i\phi)] \; ,
\end{equation}
and one can easily check that this gives a consistent relation between the canonical and grand canonical determinants. 
For a complete and exact representation of the grand canonical determinant $\det[D(\mu)]$ by the fugacity series 
(\ref{eq:fug1}) one has to set $q_{cut}$ to the maximal number $N_s^3 \times 3 \times 2$. However, already for small lattices this value cannot be 
reliably reached in numerical calculations and one is restricted to cut at smaller values, which in turn limits the values of the chemical potential that
can be reached.  In our study $q_{\text{cut}}$ is of order $O(100)$, with the actual value depending on temperature, volume and quark mass.

\begin{figure}[ht]
	\centering
	\includegraphics[width=0.49\textwidth,clip]{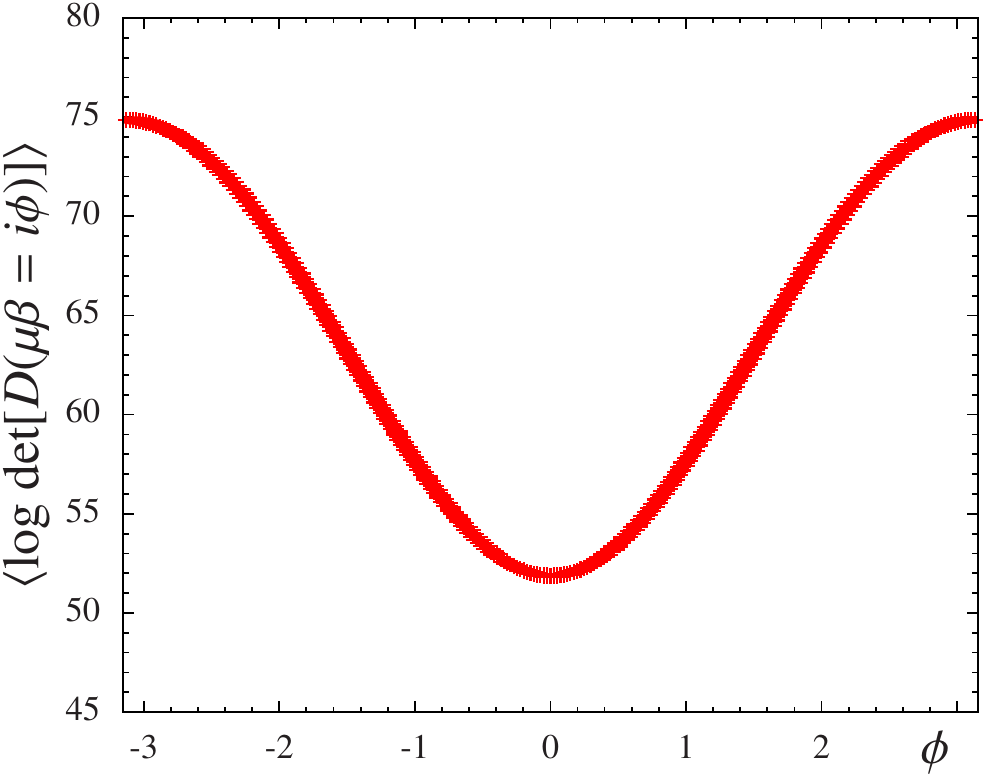}
	\hspace{0.005\textwidth}
	\includegraphics[width=0.49\textwidth,clip]{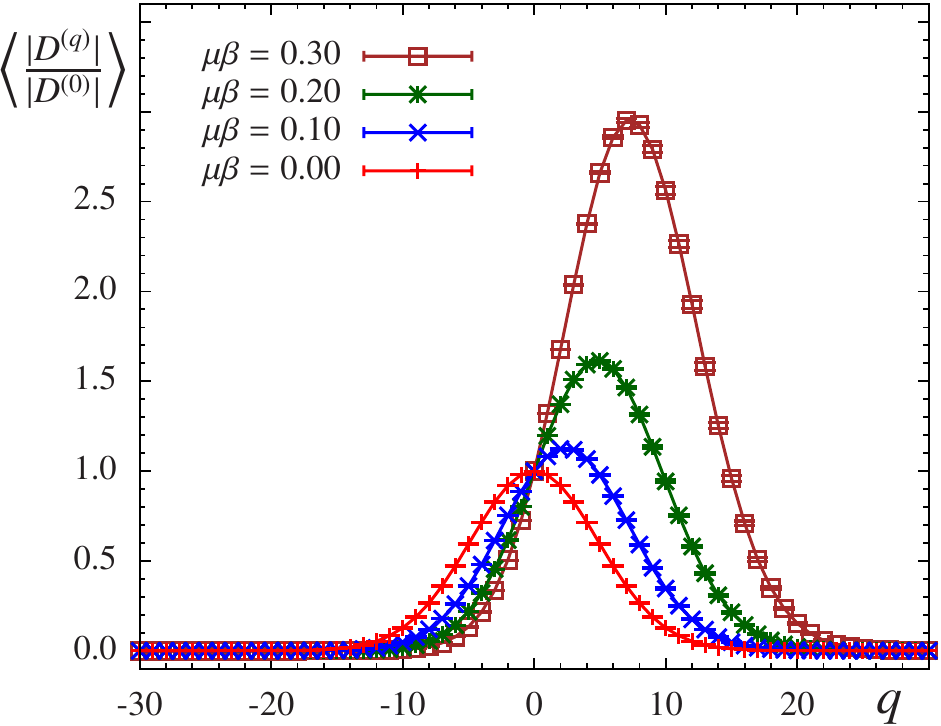}
	\caption{Lhs.: The logarithm of the modulus of the grand canonical fermion determinant $\det[D(\mu\beta = i\phi)]$ as a function of $\phi$. Rhs.:
	 The canonical determinants weighted with the corresponding fugacity factor, i.e., $D^{(q)}e^{\mu \beta q}$, as a function of the net quark number $q$ for
	 different values of the chemical potential.  All data are for Wilson fermions at $6/g^2 = 5.35$ on a $8^3 \times 4$ lattice.}
	\label{fig:qdet}
\end{figure}

\hspace*{10mm}
The canonical determinants are interesting quantities on their own and the fugacity series illustrates how a baryon chemical potential influences the system
by shifting their relative weight in the fugacity sum (\ref{eq:fug1}) for the grand canonical determinant. 
For understanding the corresponding mechanisms and the resulting numerical challenges it is helpful to take a look at the different quantities in 
Eqns.~(\ref{eq:fug1}) and (\ref{eq:fug2}). In the lhs.\ plot of Fig.~\ref{fig:qdet} we show 
the grand canonical determinant as a function of an imaginary chemical potential $\mu\beta = i\phi$. More precisely we show the logarithm of the
absolute value of $\det[D(\mu\beta = i\phi)]$. The plot illustrates that $\det[D(\mu\beta = i\phi)]$ is a quantity that varies over many orders of magnitudes
within one period of $\phi$, and the numerical challenge in the evaluation of \eqref{eq:fug2} is to compute the Fourier moments of 
$\det[D(\mu\beta = i\phi)]$ for
sufficiently high values of $q$ with high accuracy.

\hspace*{10mm}
The resulting canonical determinants $D^{(q)}$ have a Gaussian-like distribution as a function of $q$. This is illustrated by the $\mu \beta = 0.00$ data
in the rhs.\ plot of Fig.~\ref{fig:qdet}. In the fugacity sum \eqref{eq:fug1} the canonical determinants are multiplied with the corresponding 
fugacity factor and summed. These
contributions to the sum, i.e., the $D^{(q)}e^{\mu \beta q}$, for $\mu \beta = 0.10, 0.20$ and 0.30 are also shown in the rhs.\ plot of Fig.~\ref{fig:qdet}.
From this figure it is clear what effect a chemical potential has on the system: Contributions from higher quark numbers become more important as one 
cranks up the chemical potential (the curve maximum gets shifted towards higher values of $q$) and we have to calculate the canonical determinants for high 
values of $q$ if we want to get results for large values of $\mu\beta$. Furthermore we have to achieve very good accuracy in the 
evaluation of the $D^{(q)}$ because of the cancellation of the increasing factors $e^{\mu \beta q}$ against the Gaussian-like suppression of the 
 $D^{(q)}$ for large $q$. To reduce the numerical effort for this calculation we use a domain decomposition 
described in~\cite{domdecomp}.

\hspace*{10mm}
We are interested in observables related to the net quark number $q$. They can be obtained as derivatives of the grand canonical partition sum with 
respect to the chemical potential $\mu$, 
		\begin{equation}
		\frac{\chi_n^q}{T^{4-n}} = \frac{\beta^4}{V}\frac{\partial^n T \ln Z_\mu}{\partial (\mu\beta)^n} \; .
	\end{equation}
We may evaluate these derivatives directly for the fugacity series by inserting \eqref{eq:fug1} into the expression for the partition sum 
(here for staggered fermions with 2 degenerate flavors/tastes, such that $D(\mu)$ is the staggered Dirac operator with chemical potential $\mu$)
	\begin{align}
		Z_\mu &= \int D[U] \, e^{-S_g[U]}\, \,\det[D(\mu)]^{2/4}
		= \int D[U] \, e^{-S_g[U]}\, \left(\sum_{q=-q_{\text{\,cut}}}^{q_{\text{\,cut}}} e^{\,\mu \beta q}\,D^{\,(q)}\right)^{1/2} \notag\\
	&= \int D[U] \, e^{-S_g[U]} \, \det[D(\mu=0)]^{1/2} \left( \sum_{q=-q_{\text{\,cut}}}^{q_{\text{\,cut}}}  e^{\mu \beta q}\,\frac{D^{\,(q)}}{\det[D(\mu=0)]} \right)^{1/2} \; .
	\end{align}
After introducing the moments
	\begin{equation}
		M^{\,n} = \sum_{q=-q_{\text{\,cut}}}^{q_{\text{\,cut}}} e^{\,\mu \beta q}\, q^n \,\frac{D^{\,(q)}}{\det[D(\mu=0)]} \; ,
	\end{equation}
one can write the first derivative, i.e., the quark number density, as products of expectation values $\langle ... \rangle_0$ evaluated on configurations generated for $\mu=0$,
	\begin{equation}
		\frac{\chi_1^q}{T^3} = \frac{n_q}{T^3} = \frac{\;\beta^3}{V}\;\frac{1}{2}\;\frac{\langle (M^{\,0})^{-1/2} M^{\,1}\rangle_0}{\langle (M^{\,0})^{1/2}\rangle_0} \; .
	\end{equation}
The second derivative, i.e., the quark number susceptibility, depends on higher moments but can still be easily and reliably calculated,	
	\begin{equation}
	\frac{\chi_2^q}{T^2} = \frac{\;\beta^3}{V}\;\frac{1}{2}\;\left[ \frac{\langle (M^{\,0})^{-1/2} M^{\,2} \rangle_0 - \frac{1}{2}\langle (M^{\,0})^{-3/2} (M^{\,1})^2 
	\rangle_0}{\langle (M^{\,0})^{1/2} \rangle_0} - \frac{1}{2}
	\left(\frac{\langle (M^{\,0})^{-1/2} M^{\,1}\rangle_0}{\langle (M^{\,0})^{1/2}\rangle_0}\right)^2\right] \; .
	\end{equation}
Furthermore, also the 3rd and the 4th derivatives can be expressed in terms of $\mu = 0$ expectations values of moments $M^{\,n}$. 
These observables evaluated for staggered fermions are presented in Figs.~\ref{fig:stag1} and \ref{fig:stag3}. For Wilson fermions the same approach leads to similar 
equations, i.e., again
expressions that involve expectation values of moments $M^{\,n}$ at $\mu = 0$ (used in Fig.~\ref{fig:wil3}).
	
\hspace*{10mm}
Also ratios of these derivatives are interesting observables. In particular we here study
	\begin{equation}
		\frac{\chi_2^q/T^2}{n_q/T^3} \quad \mbox{and} \quad \frac{\chi_3^q/T}{\chi_2^q/T^2} \; .
	\end{equation}
These ratios can also be calculated from the hadron resonance gas (HRG)~\cite{hrg}, and we here use these results for a comparison with our data in the 
low temperature/confined regime. The HRG gives the following ratios:
	\begin{align}
		\left(\frac{\chi_2^q/T^2}{n_q/T^3}\right)_{\hspace{-0.1cm}HRG} \hspace{-0.2cm} = \hspace{0.2cm} 3 \text{sech}(3\mu\beta) \quad , \quad \left(\frac{\chi_3^q/T}{\chi_2^q/T^2}\right)_{\hspace{-0.1cm}HRG} \hspace{-0.2cm} = \hspace{0.2cm} 3 \tanh(3\mu\beta) \; .
	\end{align}
They are independent of the baryon masses and only depend on the dimensionless product $\mu\beta$. Thus they are ideal for a comparison to
lattice results.

\section{Generalized susceptibilities for Wilson fermions}
Using the approach described in Sec.~\ref{sec:two} we calculate the generalized quark susceptibilities up to the 4th order for $12^3 \times 6$ Wilson fermions with $N_f=2$ degenerate quark flavors. This ensemble consists of a minimum of $50$ (for some couplings $100$) configurations per coupling $6/g^2$ at an inverse mass parameter of $\kappa = 0.162$. The configurations were generated with the publicly available MILC code~\cite{milc}.

\begin{figure}[ht]
	\centering
	\includegraphics[width=0.485\textwidth,clip]{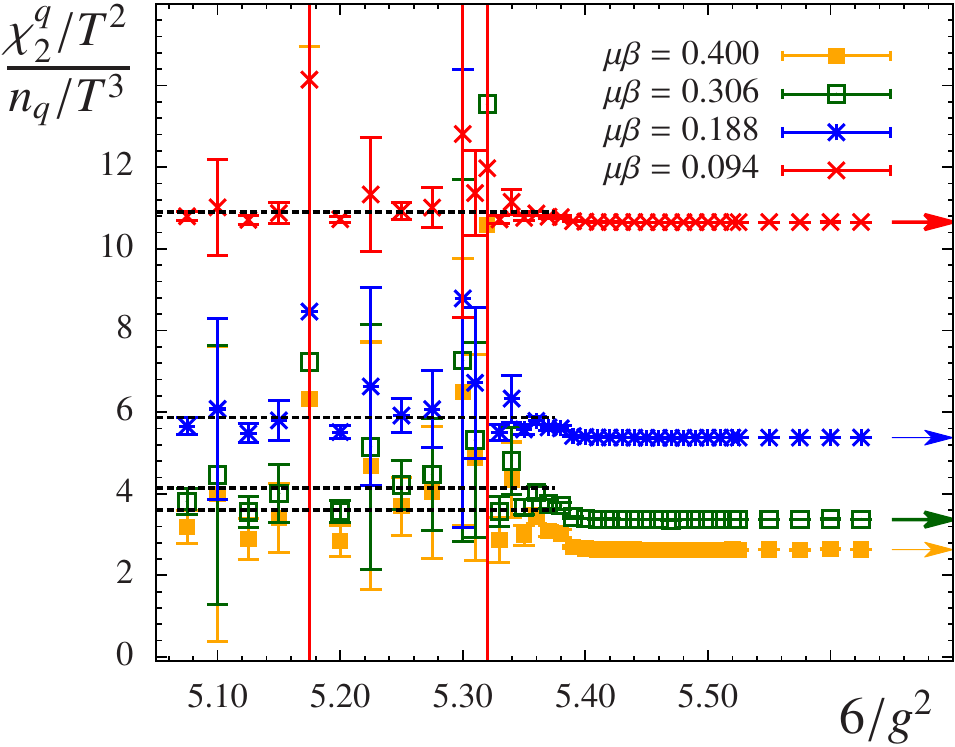}
	\hspace{0.01\textwidth}
	\includegraphics[width=0.485\textwidth,clip]{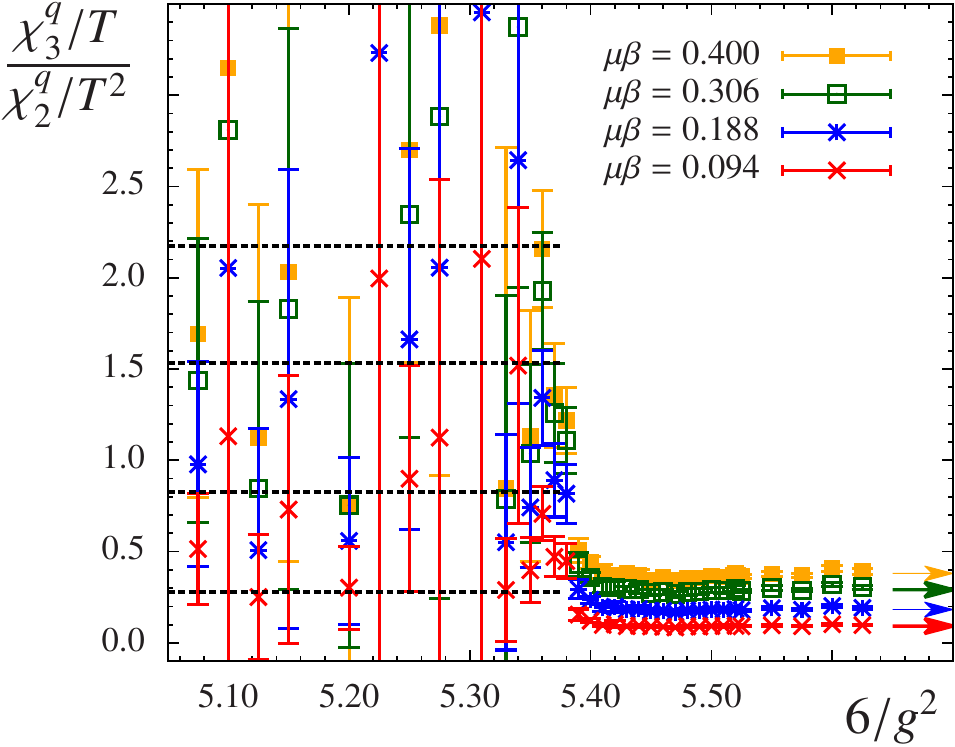}
	\caption{Ratios of derivatives as a function of the inverse coupling for different values of the chemical potential using the Wilson fermions $12^3 \times 6, \kappa=0.162$ ensemble. Dashed black lines are HRG results and the arrows on the rhs. of the plots indicate the free limit.}
	\label{fig:wil3}
\end{figure}

\hspace*{10mm}
In this preliminary presentation we only discuss the ratios of the derivatives as a function of the inverse coupling $6/g^2$, which corresponds to the temperature of the system for different values of the chemical potential, as we will use them later to compare to the staggered calculations. Fig.~\ref{fig:wil3}
demonstrates that in the confined region $(\chi_2^q/T^2)/(n_q/T^3)$ is in good agreement with the HRG. Above the crossover it rapidly approaches the free limit and stays constant already shortly above the crossover. At large inverse coupling (large temperature) the fluctuations are small. $(\chi_3^q/T)/(\chi_2^q/T^2)$ shows a similar behavior but with a more distinct change between the confined and deconfined phases. For low temperatures the statistical errors are too large to safely determine a value or decide if there is agreement with the HRG. For high temperatures, however, the ratio becomes again constant and agrees with the free limit already at an inverse coupling of $6/g^2=5.40$.

\section{Generalized susceptibilities for staggered fermions}
The calculations for staggered fermions we discuss in greater detail. We use a staggered ensemble, again with two quark flavors/tastes, with lattices of size $16^3 \times 6$. The increased lattice size is possible due to the decreased numerical effort necessary in the calculation of the staggered fermion determinant. We work at a mass parameter of $m=0.1$ with $100$ configurations per coupling.

\begin{figure}[ht]
	\centering
	\includegraphics[width=0.485\textwidth,clip]{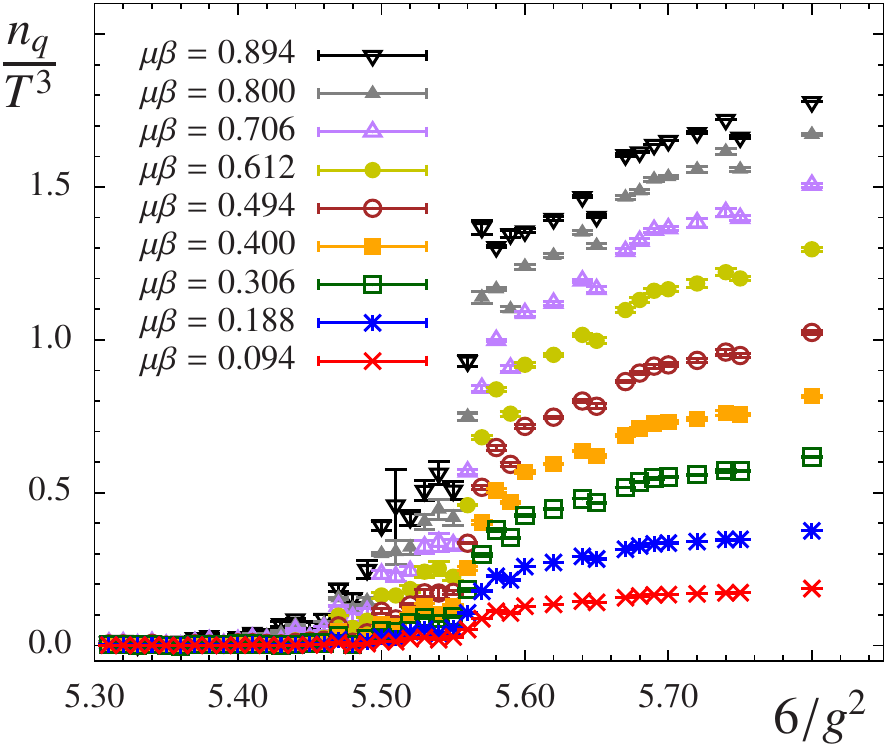}
	\hspace{0.01\textwidth}
	\includegraphics[width=0.485\textwidth,clip]{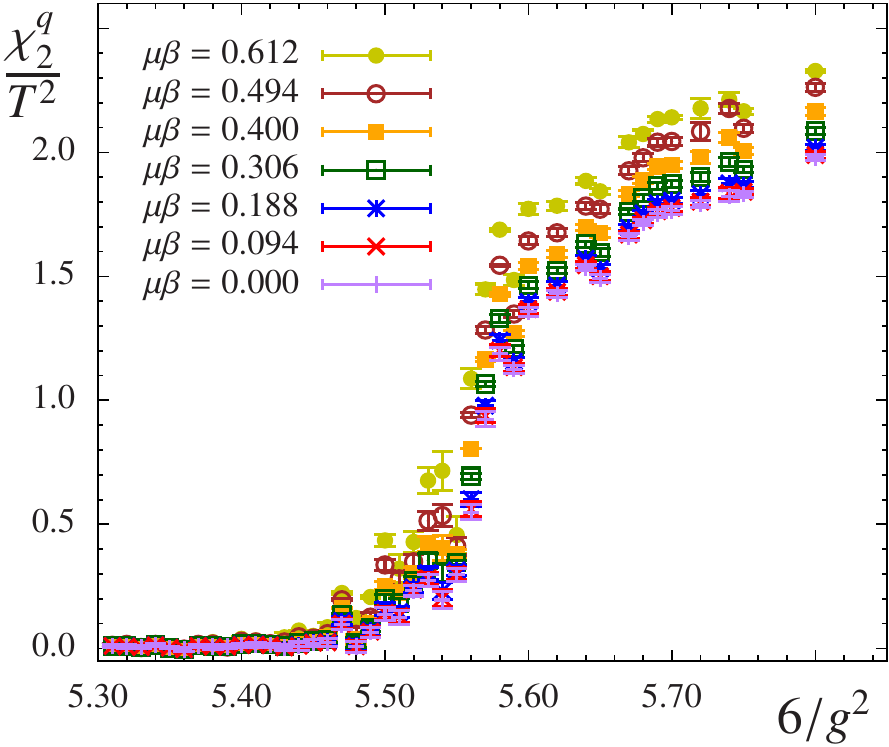}
	\caption{Quark number density (lhs.) and quark number susceptibility (rhs.) as a function of the inverse coupling for different values of chemical 
	potential using the staggered fermions $16^3 \times 6, m=0.1$ ensemble.}
	\label{fig:stag1}
\end{figure}

\hspace*{10mm}
Fig.~\ref{fig:stag1} shows the quark number density and the susceptibility as a function of the inverse coupling $6/g^2$. Around $6/g^2=5.50$ the crossover 
takes place and the small quark number density for low temperatures becomes non-zero with a strong dependence on $\mu\beta$ (for $\mu\beta=0$ it is 
always zero due to time reversal symmetry). Starting at $\mu\beta \approx 0.9$ the statistical error in the crossover region increases and the expansion starts 
to break down. For the susceptibility we observe only a small dependence on the chemical potential. The statistical errors stay small below and above the 
crossover for a chemical potential up to $\mu\beta=0.6$. The higher derivatives, which are not shown here, peak around the pseudo-critical coupling.

\begin{figure}[ht]
	\centering
	\includegraphics[width=0.485\textwidth,clip]{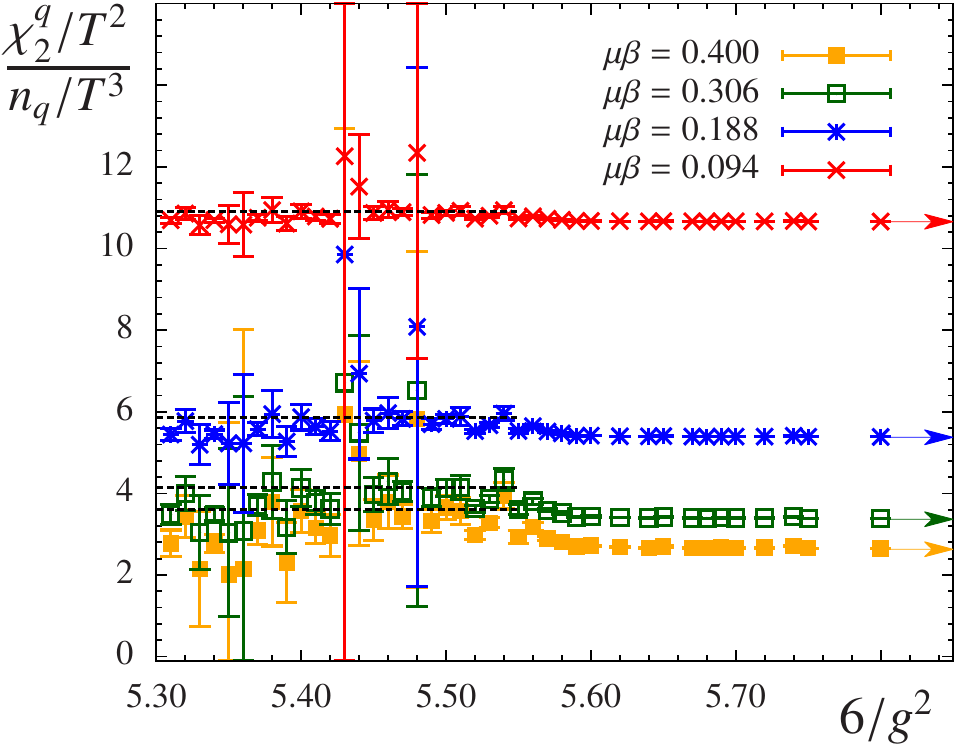}
	\hspace{0.01\textwidth}
	\includegraphics[width=0.485\textwidth,clip]{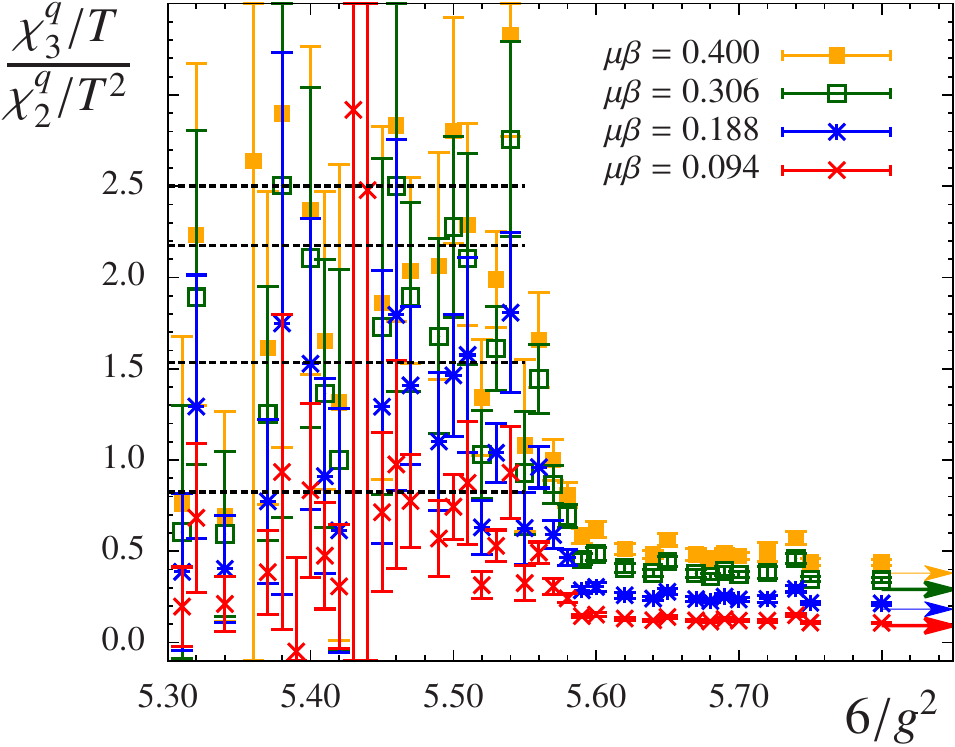}
	\caption{Ratios of derivatives as a function of the inverse coupling for different values of chemical potential using the staggered fermions $16^3 \times 6, m=0.1$ ensemble. Dashed black lines are HRG results and the arrows on the rhs.\ of the plots indicate the free limit.}
	\label{fig:stag3}
\end{figure}

\hspace*{10mm}
In Fig.~\ref{fig:stag3} ratios of the derivatives are plotted. On the lhs.\ we show $(\chi_2^q/T^2)/(n_q/T^3)$ as function of the inverse coupling,
and observe good agreement with the HRG 
results for couplings below the crossover. What we furthermore observe is good agreement with the free results at large inverse coupling (i.e., large temperatures). These free results are taken from the 
Wilson fermion calculations, which means that the ratios seem to be universal in terms of the fermion discretization used. The same can be observed for the 
second ratio $(\chi_3^q/T)/(\chi_2^q/T^2)$. Even though we cannot reliably state if there is an agreement with HRG due to the strong fluctuations in the 
confined region, we still find very good agreement with the free theory (again taken from the Wilson results). This allows us to conclude that the ratios are good 
quantities to use for comparison between different lattice discretizations as their dependence on the discretization is weak. By determining the scale of both 
ensembles, a direct comparison would be possible and should be performed before drawing final conclusions.

\section{Conclusions and final remarks}
In this project we explore the fugacity expansion as an approach to extract quark number susceptibilities at finite $\mu$ from lattice QCD calculations. 
We presented results 
for Wilson and staggered ensembles and compare their generalized susceptibilities and ratios. We found good qualitative agreement between the different 
discretizations and also good agreement with HRG calculations for the ratios in the confined phase. Furthermore, the ratios from the different ensembles 
approach the same free limit in the deconfined phase and one may conclude, that the ratios are not only interesting for comparison with experiment, but are 
also good observables to cross check different lattice results.

\hspace*{10mm}
With the fugacity expansion one has another method at hand to perform analytic continuations to finite $\mu$. It has the specific advantage, that once the
canonical determinants $D^{(q)}$  are known (although the numerical effort for their evaluation is high), all quark number related observables can be accessed very easily.

\vspace{0.2cm}
{\bf Acknowledgments:}
We thank T.~Kloiber, C.B.~Lang, A.~Schmidt, K.~Splittorff and J.~Verbaarschot for fruitful discussions. H.-P.~Schadler is 
funded by the FWF DK W1203 ``{\sl Hadrons in Vacuum, Nuclei and Stars}''. Furthermore this project is partly  supported by DFG TR55, 
``Hadron Properties from Lattice QCD'' and by the Austrian Science Fund FWF Grant. Nr. I 1452-N27.

\end{document}